\documentclass[12pt]{article}

\usepackage{graphicx}
\usepackage{amssymb}
\usepackage{citesort}
\usepackage{url}
\usepackage{amsfonts,amsmath,amscd,mathrsfs,comment}

\usepackage{color}

\def\pa{\partial}

\begin{document}
\title{Can  Bohmian mechanics be made relativistic?}

\author{
Detlef D\"urr\footnote{Mathematisches Institut, Ludwig-Maximilians-Universit\"at, 
	Theresienstra{\ss}e.~39, 80333~M\"unchen, Germany. E-mail: duerr@mathematik.uni-muenchen.de},
Sheldon Goldstein\footnote{Departments of Mathematics, Physics and
     Philosophy, Rutgers University, Hill Center,  
     110 Frelinghuysen Road, Piscataway, NJ 08854-8019, USA.
     E-mail: oldstein@math.rutgers.edu},
Travis Norsen\footnote{Smith College, Northampton, MA 01060, USA. E-mail: tnorsen@smith.edu},\\
Ward Struyve\footnote{Departments of Mathematics and Philosophy,
     Rutgers University, Hill Center,  
     110 Frelinghuysen Road, Piscataway, NJ 08854-8019, USA.
     E-mail: wstruyve@math.rutgers.edu},
and Nino Zangh\`\i\footnote{Dipartimento di Fisica dell'Universit\`a
     di Genova and INFN sezione di Genova, Via Dodecaneso 33, 16146
     Genova, Italy. E-mail: zanghi@ge.infn.it}
}

\maketitle

\begin{abstract}
\noindent
In relativistic space-time, Bohmian theories can be formulated by
introducing a privileged foliation of space-time. The introduction of
such a foliation -- as extra absolute space-time structure -- would seem
to imply a clear violation of Lorentz invariance, and thus a conflict
with fundamental relativity. Here, we consider the possibility that,
instead of positing it as extra structure, the required foliation
could be covariantly determined by the wave function. We argue that this
allows for the formulation of Bohmian theories that seem to qualify as
fundamentally Lorentz invariant. We conclude with some discussion of
whether or not they might also qualify as fundamentally relativistic.
\end{abstract}

\renewcommand{\baselinestretch}{1.1}
\bibliographystyle{unsrt}

\section{Introduction}
\label{sec1}

Bohmian mechanics, also known as the de Broglie-Bohm pilot-wave
theory, is a version of quantum mechanics in which the notions of 
observation, measurement, and the macroscopic world play no
fundamental role \cite{bohm93,holland93b,duerr09,duerr12}.  For the case of an $N$-particle universe of
spinless non-relativistic particles, the theory posits a universal
wave function obeying the usual Schr\"odinger wave equation
\begin{equation}
i \hbar \frac{\partial \Psi_t}{\partial t} = -\sum_{k=1}^{N}
\frac{\hbar^2}{2 m_k} \nabla_k^2 \Psi_t + V \Psi_t
\label{eq-nrsch}
\end{equation}
as well as particles with definite positions evolving according to the
guidance formula
\begin{equation}
\frac{dX_k(t)}{dt} = \frac{\hbar}{m_k} \text{Im} \frac{\Psi^*_t \nabla_k
  \Psi_t}{\Psi^*_t \Psi_t} \big|_{X_1(t), X_2(t), ... , X_N(t)}.
\label{eq-nrguidance}
\end{equation}
Here $\Psi_t$, the universal wave function, is a 
complex-valued function on the $3N$-dimensional configuration space of
the $N$ particles.  (For particles with spin, one need only consider
$\Psi_t$ as instead being the appropriate $N$-particle spinor, obeying
instead of Equation \eqref{eq-nrsch} the appropriate wave equation,
and then
interpret the numerator and denominator of the right hand side of
Equation \eqref{eq-nrguidance} as involving the appropriate spinor
inner products.)

One common objection to Bohmian mechanics is that it is  incompatible with, and cannot be made
compatible with, relativity.  Of course, the theory sketched above is
a version of non-relativistic quantum mechanics, so one expects
incompatibility with relativity.  The point, though, is that the
non-relativistic theory contains a certain feature (suggesting
incompatibility with relativity) that persists even when, for example,
Equation \eqref{eq-nrsch} is replaced with a relativistic wave
equation (such as the Dirac equation):  the guidance formula, Equation
\eqref{eq-nrguidance}, has, at any time, the velocity of each particle being defined
in terms of the wave function and its spatial derivatives \emph{evaluated
  at the actual configuration point of the entire $N$-particle system
  at that time}.  The velocity of a given
particle thus depends, in general, on the instantaneous positions of all
the other particles.  

Bohmian mechanics is, therefore, a non-local theory in a precise sense
articulated especially by Bell.  Indeed, the non-locality just
described is precisely the manifestation, in this theory, of the sort
of non-locality Bell proved must be present in any theory sharing the
empirical predictions of ordinary quantum theory \cite{bell87a,goldstein11}.  

Unlike certain
other ``quantum theories without observers'' (for example, those for which outcome independence but not parameter independence is violated)\footnote{For some results concerning relativistic collapse theories see \cite{tumulka06a,bedingham11}.},
however, the Bohmian type of non-locality has seemed particularly
difficult to reconcile with fundamental relativity.  Leaving aside the
trivial case of a single particle, the usual guidance formula for Bohmian
theories involving appropriately relativistic wave equations for
$\Psi$ requires something like a privileged Lorentz frame or family of simultaneity surfaces. 

Many proponents of Bohmian ideas have thus become resigned to the 
notion that relativistic Bohmian theories will be relativistic only at
the relatively superficial level of empirical predictions:  the
theories will make relativistically good predictions (including, for
example, the correct kind of prediction for the Michelson-Morley
experiment) but will involve something like a hidden, empirically 
undetectable, notion of absolute simultaneity.  Such theories,
it is usually conceded, are relativistic only in the sense that the
Lorentz-Fitzgerald ether theory (considered here as an interpretation
of classical electrodynamics) is relativistic -- namely, they are
\emph{not} relativistic, not in a \emph{serious} or \emph{fundamental}
sense.  

The aim of the present paper is to question this perspective by
suggesting a rather general strategy for making Bohmian theories
compatible with fundamental relativity. In Bohmian theories the
dynamics of the particles (or fields) is defined in terms of
structures extracted from the wave function. The strategy proposed
here involves extracting from the wave function also a foliation of
space-time into space-like hypersurfaces, which foliation is used to define a Bohmian dynamics in a manner similar to the way equal-time hyperplanes are used to define the usual Bohmian dynamics. We show
how this extraction can itself be Lorentz invariant in an appropriate
sense, and argue that virtually any relativistic quantum
theory, Bohmian or otherwise, will thus already contain a special
space-time foliation, buried in the structure of the wave function.
This makes it difficult to imagine how one could question the
``seriously relativistic'' character of the Bohmian theories to be
described, without simultaneously denying that any theory in which
something like a universal wave function plays a fundamental role can be a
candidate for serious compatibility with relativity.

Bohmian formulations that employ a foliation extracted from the wave function (which were considered before in \cite{duerr90,duerr99}) reproduce the standard quantum predictions (as we will explain in section \ref{sec6}). A variety of alternative approaches yield a covariant Bohmian dynamics but seem to disagree with the standard quantum predictions. Here is a brief overview. First, one can employ a foliation that may depend also on the actual configuration \cite{bell84,duerr90,horton04}. Second, rather than using a foliation, there also have been attempts to use only the relativistic light-cone structure instead \cite{squires93,goldstein03}. Finally, there are approaches that do not make use of surfaces but instead introduce a ``synchronization'' of particle trajectories \cite{berndl96a,horton01,dewdney02,nikolic05b}. For more details, see \cite{tumulka07}.

The argument is developed as follows in the subsequent sections of the
paper.  In Section \ref{sec2} we briefly review the formulation of
Bohmian theories in relativistic space-time, giving simple examples
involving both particles and fields.  Section \ref{sec3} describes the
construction of geometrical structures in space-time using the tools
available from quantum field
theories, while Section \ref{sec4} discusses the meaning of serious or
fundamental Lorentz invariance and articulates in particular what it
would mean for a foliation to be extracted from the wave function in a
relativistically covariant manner.  Section \ref{sec5}
sketches a possible way to generalize one of the suggestions made in
Section \ref{sec3} which does not require a foliation, while Section \ref{sec6} takes up the question of
the relationship between the statistical predictions of the theories
proposed here to those of ordinary relativistic quantum theories.
Finally, in Section \ref{sec7}, we reflect on the question posed in the paper's title.

\section{Bohmian theories in relativistic space-time}
\label{sec2}

In the case of a single particle ($N=1$), the non-relativistic Bohmian
theory sketched in the introduction is a local theory:  the
configuration space is isomorphic to physical space, and the guidance
formula, Eq.~\eqref{eq-nrguidance}, can thus be understood as determining
the particle's velocity (at each instant) in terms of the structure of 
the wave function around the location of the particle.  In the more
general case ($N > 1$) however -- and here assuming the general case
of a non-factorizable, entangled wave function --  
the non-locality of the theory
(necessary for its ability to correctly predict violations of Bell's
inequality) becomes manifest:  each particle's velocity is (at each
instant) defined in terms of the structure of the wave function around the
actual configuration point of the whole $N$-particle system.  The 
motion of each particle therefore depends 
on the instantaneous positions of all the other (distant) particles. This non-locality is the root of the difficulty involved in reconciling Bohmian approaches to quantum theory with serious,
fundamental Lorentz invariance.  

To further explain the difficulty,
and to explain the general structure of Bohmian theories in the
context of relativistic space-time, let us consider the generalization
of (non-relativistic) Bohmian mechanics to relativistic 
particles and to quantum fields obeying a relativistic wave equation.

Let us start by outlining the general structure of a Bohmian formulation for $N$ particles in Minkowski space-time $M$. A possible history of the particles arising from their motion corresponds to an $N$-path $X=(X_1,\dots,X_N)$, where $X_k=X_k(s)$ is the space-time trajectory of the $k^{\textrm{th}}$ particle. The law of motion for the particles, which defines the dynamically allowed $N$-paths, is expressed in terms of a wave function $\Psi$ and a foliation $\mathscr{F}$ of space-time into space-like hypersurfaces (not necessarily hyperplanes).

The role of the foliation is to provide the notion of a configuration
to be inserted in an evolution equation like
\eqref{eq-nrguidance}. Such a configuration is given by the crossings
$(X_1^\Sigma, ... , X_N^\Sigma)$ by the space-time paths of the
particles of any leaf $\Sigma$ of the foliation. The evolution law can
be formulated by demanding that the (unit) tangent vector ${\dot
  X}_k^\Sigma$ to the trajectory of the $k^{\textrm{th}}$ particle at
the point through which it crosses $\Sigma$ be parallel to a vector
$v^{\mathscr{F}, \Psi{}}_k$ depending on the configuration on
$\Sigma$, as well as on the wave function $\Psi$ and possibly the
foliation ${\mathscr{F}}$. The law can then be written in the form
\begin{equation} 
{\dot X}_k^\Sigma \propto  v^{\mathscr{F}, \Psi{}}_k (X_1^\Sigma, ... , X_N^\Sigma)\,,
\label{particleguidance}
\end{equation}
where $v^{\mathscr{F}, \Psi}_k$ is a function on $M^N$. (In fact it would be sufficient if $v^{\mathscr{F}, \Psi{}}_k$ were defined only for $N$-tuples of space-time points on a common leaf of the foliation). ``$\propto$'' means proportional to, with a positive (perhaps $x$-dependent) proportionality constant. Different such proportionality ``constants'' will merely correspond to different parameterizations of the same trajectories. 

As an example, we consider free (positive-energy) Dirac particles.  It is convenient to use the multi-time wave function $\Psi = \Psi(x_1, ..., x_N)$, where the $x_k \in M$ are space-time points, and which takes values in the $N$-particle spin space $({\mathbb C}^4)^{\otimes N}$ \cite{duerr99}. It satisfies the $N$ Dirac equations{\footnote{From now on, we assume $\hbar=c=1$.}}
\begin{equation}
i  \gamma^\mu_k \partial_{k,\mu} \Psi  -  m_k \Psi = 0
\label{multitimedirac}
\end{equation}
for $k = 1, ..., N$ (where as usual the summation convention over $\mu$ is assumed).  Here, $\gamma^\mu_k = I \otimes \cdots \otimes I \otimes \gamma^\mu \otimes I \otimes \cdots \otimes I$ with the Dirac matrix $\gamma^\mu$ at the $k$-th of the $N$ places. 

The velocity fields $v^{\mathscr{F}, \Psi{}}_k$ that appear in the dynamical equation \eqref{particleguidance} for the particles are
\begin{equation}
\left[v^{\mathscr{F}, \Psi{}}_k \right]^{\mu_{k}}(x_1, \ldots, x_N)= J^{\mu_1 \ldots \mu_N} ( x_1, \ldots, x_N)  n^{\mathscr{F}}_{\mu_1}(x_1) \ldots \widehat{n^{\mathscr{F}}_{\mu_k}(x_k)}\ldots   n^{\mathscr{F}}_{\mu_N}(x_N) ,
\label{EQU1}\end{equation}
where $\;\widehat{}\;$ indicates that the term should be omitted, $n^{\mathscr{F}}$ is the unit future-directed normal vector field to the foliation ${\mathscr{F}}$ and 
\begin{equation}  
J^{\mu_1 \ldots \mu_N} (x_1, \dots, x_N) = \overline{\Psi} ( x_1, \ldots, x_N) \gamma_1^{\mu_1} \ldots \gamma_N^{\mu_N} \Psi ( x_1 , \ldots, x_N) .
\label{eqJ}
\end{equation}
Note that although we have written down a theory of
free particles, their motions will still be coupled if the wave
function is entangled.  The theory thus includes precisely (but only)
the sort of ``interactions'' that make it so difficult to reconcile
Bohmian mechanics with fundamental relativity.  

In the special case of a single particle, the wave function satisfies the Dirac equation and the particle law reduces to
\begin{equation}
\frac{d X(s)}{ds} \propto v^\Psi(X(s)) = J(X(s)) = \overline{\Psi} (X(s) ) \gamma \Psi (X(s)) .
\label{eq-1pdiracguidance}
\end{equation}
So in this case the law does not depend on the foliation and is fully Lorentz invariant. 

There also exist Bohmian theories with an actual field configuration $\varphi(x)$ on space-time instead of particle positions (at least for bosonic quantum fields \cite{struyve10}). As in the case of particles, the evolution equation of the field in a relativistic version should depend on the wave function $\Psi$ and foliation $\mathscr{F}$ \cite{duerr90,horton04}. It should be of the form
\begin{equation}
\frac{d \varphi (x)}{d\tau} = F^{\mathscr{F},\Psi}(x,\varphi|_{\Sigma_x}) ,
\label{fieldguidance}
\end{equation}
where $d \varphi /d\tau$ is the directional derivative at $x$ along the normal to the leaf $\Sigma_x$ of the foliation that contains $x$, i.e., $d \varphi /d\tau = n^{\mathscr{F}} \cdot \partial \varphi$. $\varphi|_{\Sigma_x}$ is the restriction of the field configuration to the hypersurface $\Sigma_x$. 

For example, for a real massless scalar field $\varphi(x)$, we have
\begin{equation}
F^{\mathscr{F},\Psi} = \text{Im} \left( \frac{1}{\Psi_{\Sigma_x}} \frac{\delta \Psi_{\Sigma_x}}{\delta \varphi_{\Sigma_x}(x)} \right)\Big|_{\varphi|_{\Sigma_x}},
\label{f}
\end{equation} 
where for each space-like hypersurface $\Sigma$ the wave function
$\Psi_\Sigma$ is a functional $\Psi_\Sigma(\varphi_\Sigma)$ of fields
$\varphi_\Sigma$ on $\Sigma$. The wave functional can be defined as
$\langle \varphi_\Sigma | \Psi \rangle$, where $|\Psi \rangle$ is the
state vector in the Heisenberg picture and where
$|\varphi_\Sigma\rangle$ is defined by $ \varphi(x)
|\varphi_\Sigma\rangle = \varphi_\Sigma(x) |\varphi_\Sigma\rangle$,
for points $x$ on $\Sigma$, with $\varphi(x)$ the Heisenberg field
operator.  (Since the field operators commute at space-like separated points, $|\varphi_\Sigma\rangle$ is well-defined as the simultaneous eigenstate of the field operator at different points on $\Sigma$.) $\delta /\delta \varphi_\Sigma$ is the functional derivative (which for a functional $G$ is defined by $\int_\Sigma d\sigma \frac{\delta G(\varphi_\Sigma)}{\delta \varphi_\Sigma}f = \frac{d}{d\epsilon} G(\varphi_\Sigma + \epsilon f)\big|_{\epsilon = 0}$). Similarly as in the case of (entangled) Dirac particles, the field evolution will depend on the choice of foliation.

The point of sketching the two examples above was to
indicate the general structure of the Bohmian versions of relativistic
quantum theories.
In general, such theories will involve, in addition to the usual quantum
mechanical wave function of the appropriate sort, two pieces of
additional structure:  the ``local beables'' or ``primitive ontology''
(which, in the above examples, were the actual configuration of $N$
particles and the scalar field, respectively\footnote{Particle and field configurations do not exhaust the possibilities
  for primitive ontology in Bohmian theories.  Other sorts of objects,
  like strings, may be considered in this role, and more complex
  theories  may combine several of the elements
  already mentioned.  It should also be emphasized that a Bohmian
  version of (what is usually called) a relativistic field theory 
  need not use field configurations as the local beables.  It is
  possible, for example, to consider a field theory with particle
  positions as the local beables \cite{duerr04b,colin07}.})
and a foliation of space-time.

It has been 
suggested that such a theory -- involving such a privileged
foliation of space-time -- 
might nevertheless be considered a candidate for ``serious'' compatibility with
fundamental relativity if the foliation is not simply posited as a
novel piece of absolute space-time structure, but is instead regarded
as a dynamical object, itself obeying a Lorentz invariant law. Examples of such a law are $\pa_\nu n_\mu =0$ \cite{duerr99} or $\pa_\nu n_\mu - \pa_\mu n_\nu =0$ \cite{tumulka07}, where $n$ is the future-directed normal vector field to the foliation. (For the first example it means that the leaves of the foliation are hyperplanes, for the second it means that the time-like distance from a point on a leaf to another leaf is constant along the leaf.) Such models, though, seem to have a strong un-relativistic flavor even if they are in some sense fundamentally Lorentz invariant.  The Lorentz invariance seems, somehow, unserious.

A different idea, which can perhaps be taken more seriously as
providing genuinely relativistic theories, is to extract the needed
foliation $\mathscr{F}$ of space-time from the wave function itself 
\begin{equation}
\Psi \rightarrow \mathscr{F} = \mathscr{F}(\Psi)
\end{equation}
instead of positing it as new structure. In this case, the velocity
field of Equation \eqref{particleguidance} becomes $v^{\mathscr{F},
  \Psi{}}_k=v^{ \Psi{}}_k$ (and similarly for $F$ of Equation
\eqref{fieldguidance}).   This is the idea we will 
develop in subsequent sections.

\section{{Geometrical  structures on space-time and quantum  field theory}}
\label{sec3}

Given that the familiar representation of the wave function $\Psi$ 
lives on configuration space (or a field space) 
it is not immediately clear how it can provide a
geometrical structure on space-time itself, in particular the foliation
$\mathscr{F}= \mathscr{F}(\Psi)$.  So the first question we must
address is:  How can one extract geometrical structures on space-time 
from the wave function?
Here we develop the thought that, as soon as one considers quantum
field theory, natural possibilities suggest themselves.

Conventional formulations of quantum field theory (QFT) are in terms of fields operators $\phi(x)$ transforming covariantly under the action of the Poincar\'e group,
\begin{align}
U_{(a,A)} \phi(x) U_{(a,A)}^{-1} = D_A^{-1} \phi (Ax+a)\,,
\label{eq:qfinvariance}\end{align}
where $a$ is a translation and $A$ is a Lorentz transformation; $(a, A) \mapsto U_{(a,A)}$ is a unitary (projective) representation of the Poincar\'e group  on the Hilbert space $\mathscr{H}$ of quantum states (generated by polynomials of the fields acting on the vacuum state $|0\rangle$);   $D$ is a projective representation of the Lorentz group, e.g., for a Dirac field, $D_A$ are 4$\times$4 matrices suitably expressed by means of the gamma matrices. 

Thus, however abstract the wave function might be, the
standard local Heisenberg field operators of QFT can be used to define structures in
space-time. For example, with $j^\mu(x)$, $s^{\mu\nu}(x)$ and $t^{\mu\nu}(x)$ respectively the charge current, spin tensor and energy-momentum tensor operator, one can define the following tensorial objects for a wave function $\Psi$:
\begin{equation} 
J^{\mu}(x)= \langle \Psi{} | j^\mu(x) | \Psi{} \rangle, \qquad S^{\mu \nu}(x)  = \langle \Psi | s^{\mu\nu}(x) | \Psi{} \rangle  , \qquad T^{\mu \nu}(x) = \langle \Psi | t^{\mu\nu}(x) | \Psi{} \rangle \,.
\label{tensors}
\end{equation}
In the case of the free Dirac field $\psi(x)$, we have{\footnote{The colons denote normal ordering (i.e., annihilation operators that appear in the plane wave expansion of the field operators are moved to the right and every reordering induces a change of sign, due to the anti-commuting character of the Dirac fields). We further set aside possible issues that may arise in making these tensors  well-defined in the context of theories with interactions.}}
\begin{equation} 
j^{\mu}(x)=  : \bar{\psi} (x)\gamma^\mu \psi (x):  ,\qquad s^{\mu\nu}(x) = :\bar{\psi} (x)  \frac{i}{2} \left[
  \gamma^\mu, \gamma^\nu \right]   \psi(x):  ,
\nonumber
\end{equation}
\begin{equation} 
t^{\mu\nu}(x) = : 
\bar{\psi}(x) \frac{i}{2} \left(
  {\overleftrightarrow{\partial^\mu}}\gamma^\nu +
  {\overleftrightarrow{\partial^\nu}}\gamma^\mu \right) \psi(x):,
\end{equation} 
where
${\overleftrightarrow{\partial^\mu}} = \frac{1}{2} \left( 
{\overleftarrow{\partial^\mu}} + {\overrightarrow{\partial^\mu}} \right)$.
Such formulas for geometrical (tensor) fields are of course 
naturally covariant.

A QFT tensor that is a vector field,  such as
$J^\mu$ above,  is already very close to what one needs to define
a foliation of space-time, namely the foliation that is orthogonal to
the vector field.  
Of course, to have a foliation of space-time into
space-like hypersurfaces, the vector should be time-like -- a property
indeed possessed by $J^\mu$ in the case of Dirac QFT.  However, the foliation will not be
well-defined  unless the vector field is integrable \cite{hicks65} -- a property
\emph{not} necessarily possessed by $J^\mu$.  One may take its
integrable part \cite{duerr99}, but this may then no longer be time-like.  So
some massaging may be necessary to extract a foliation in this way. We will not take up the needed massaging here. Instead we will explore, in Section \ref{sec5}, an alternative -- and more general -- way to formulate a Bohmian dynamics which employs instead of a foliation, merely a time-like vector field, which may not be integrable. In addition, in the next section we will consider an unproblematic and simple way of defining a foliation in terms of the energy-momentum tensor $T^{\mu \nu}$.

Apart from geometrical structures of the type just considered, which are structures on space-time, we also have geometrical structures on the Cartesian product of $N$ copies of space-time. Considering again the free Dirac QFT, an example is the tensor
\begin{equation}
J^{\mu_1 \dots \mu_N}(x_1, \ldots , x_N) =  \langle \Psi | : \frac{1}{N!} \bar{\psi}(x_1) \gamma^{\mu_1}\psi(x_1) \ldots  \bar{\psi}(x_N) \gamma^{\mu_N}\psi(x_N) : |\Psi \rangle .
\label{multicurrent}
\end{equation}
This tensor defines, via \eqref{particleguidance} and \eqref{EQU1}, the dynamics of the Dirac particles since it is the same as  the one given in \eqref{eqJ} with  multi-time wave function 
given by \begin{equation}\label{wf}
\Psi(x_1,\dots,x_N) = \frac{1}{\sqrt{N!}} \langle 0| \psi(x_1) \dots \psi(x_N) |\Psi\rangle .
\end{equation}
(As the Dirac field operator $\psi(x)$ satisfies the free Dirac equation, it is immediately clear that \eqref{wf} satisfies the equations \eqref{multitimedirac}.)

\section{Lorentz invariance}
\label{sec4}
The next question we want to address is that of the Lorentz invariance
of a Bohmian theory constructed along the lines sketched above. 

In general, a physical theory posits some dynamical objects
and some law(s) governing their behavior.  For a given slate of
posited objects, there will be a set $\mathscr{X}$ of their
kinematically possible histories.  The law can then be understood as the subset $\mathscr{L} \subset
\mathscr{X}$ of histories which
are dynamically allowed by the theory. 
The invariance of a theory under a certain symmetry group
means that the set $\mathscr{L}$ is invariant under
the action of the group. For example, for the theory of the
classical massless Klein-Gordon field, $\mathscr{X}$ contains the smooth functions $\phi(x)$ on space-time and the subset $\mathscr{L}$ contains those functions that satisfy $\pa_\mu \pa^\mu \phi =0$.

A Bohmian theory generally includes in its ontology both a wave function $\Psi$ and some
objects -- the local beables or the primitive ontology --
which could for example correspond to particles or fields.  The wave
function $\Psi$ satisfies an appropriate wave equation, while the
local beables obey a guidance equation involving $\Psi$. We shall denote by $X$ a kinematically possible history of the local beables.  

Under a group $G$ of space-time transformations (such as the Lorentz
or Poincar\'e groups) $X$ will transform  
according to $X' = \Lambda_g X$, where $\Lambda_g$ is the natural
action of an element $g \in G$ on $X$.  The 
wave function will transform according to an appropriate 
representation $U_g$ of $G$:  $\Psi' = U_g \Psi$.  The invariance of the theory then means
that
\begin{equation}
\left( X', \Psi' \right) \in {\mathscr{L}} \;  \Longleftrightarrow \; \left( X,
  \Psi \right) \in {\mathscr{L}} .
\end{equation}

Alternatively, writing $\mathscr{L}^\Psi$ for the set of possible
histories $X$ for a given $\Psi$, then the invariance of the theory
under the group $G$ means that
\begin{equation}
X' \in \mathscr{L}^{\Psi'}
\;  \Longleftrightarrow \; X \in \mathscr{L}^\Psi.
\end{equation}
Or equivalently, the diagram 
\begin{equation}
 \CD
\Psi{}    @>>>    {\mathscr{L}}^\Psi{}\\
@VU_gVV         @VV \Lambda_{g} V\\
\Psi{}'  @>>>    {\mathscr{L}}^{\Psi{}'}
\endCD
\end{equation}
is commutative.

If the set  $\mathscr{L}^\Psi$ is characterized in terms of
geometrical structures on space-time, as we have seen above (for
example, for one Dirac particle, the trajectories $X$ are integral
curves of the vector field $  \overline{\Psi} ( x) \gamma  \Psi (
x)$), i.e. if  
\begin{equation}
\mathscr{L}^\Psi \longleftrightarrow (  \underbrace{\ldots,
  \Gamma^\Psi, \ldots}_{\text{various structures}}  ) 
\end{equation}
we have a similar statement for each of the structures
\begin{equation}
 \CD
\Psi{}    @>>>    {\Gamma}^\Psi{}\\
@VU_gVV         @VV \Lambda_g V\\
\Psi{}'  @>>>    {\Gamma}^{\Psi{}'}
\endCD\label{eq2}
\end{equation}
where $\Lambda_g $ is the natural action of $g$ on the
geometrical structures.\footnote{ The commutative diagram must hold when
    ${\Gamma}^\Psi{}$ is uniquely determined by
    $\mathscr{L}^\Psi$. But if different 
${\Gamma}^\Psi{}$s, say ${\Gamma}^\Psi_1$  and ${\Gamma}^\Psi_2$, both
correspond to the same  $\mathscr{L}^\Psi$, so that we may write
${\Gamma}^\Psi_1 \sim {\Gamma}^\Psi_2$ and regard them as equivalent,
then instead of $$\Lambda_g {\Gamma}^\Psi =  {\Gamma}^{\Psi'}$$
(equation to obtain \eqref{eq2}), we need only require
$$\Lambda_g {\Gamma}^\Psi \sim  {\Gamma}^{\Psi'}\,,$$ 
i.e., that the two structures define the same law. In this case, of
course, we have the commutative diagram for the equivalence classes of
the $ {\Gamma}^\Psi$.  For example, one could imagine a Bohmian theory for one particle where the velocity field $v^\Psi$ transforms as $\Lambda_g v^\Psi = \alpha v^{\Psi'}$, where $\alpha$ is a non-zero constant. While the velocity field hence does not transform covariantly, the Bohmian guidance law $dX(\tau)/d\tau \propto v^\Psi (X(\tau))$ remains covariant.} We then say that ${\Gamma}^\Psi$ is covariant. 

Examples of covariant geometrical structures are given by the tensors \eqref{tensors}. Consider, for example, the vector field $J^\Psi$ in the case of Dirac QFT. The natural action $\Lambda_g$ on $J^\Psi$ is 
\begin{equation} 
(\Lambda_g J^\Psi) (x) = L_g \left[ J^\Psi (g^{-1} x) \right],
\end{equation}
where $L_g$ is the natural action of $g$ on vectors at $g^{-1} x$ (i.e., in the tangent space at $g^{-1} x$). On the other hand, from the basic transformation laws 
\eqref{eq:qfinvariance} of the  quantum fields and the covariance property of the gamma matrices,
\begin{equation}D^{-1}_g \gamma D_g = L_g \gamma\,, \end{equation}
one obtains
$$  L_g \left[ J^\Psi (g^{-1} x) \right]=   \langle \Psi{} | : \bar{\psi{}} (g^{-1} x) L_g \gamma 
  \psi{} (g^{-1} x)): | \Psi{} \rangle 
=  J^{U_g\Psi}\,,$$
which establishes the covariance of $J^\Psi$. The covariance of the other tensors in \eqref{tensors} follows similarly. 

For our purposes, the most relevant geometrical structure is 
the foliation $
{\mathscr{F}}^\Psi$. Its covariance is expressed by the commutative  diagram
\begin{equation}
 \CD
\Psi{}    @>>>    {\mathscr{F}}^\Psi{}\\
@VU_gVV         @VV \Lambda_g V\\
\Psi{}'  @>>>    {\mathscr{F}}^{\Psi{}'} .
\endCD
\end{equation}
Here the natural action $ \Lambda_g$ on the foliation is the action of
$g$ on any leaf $\Sigma$ of the foliation.  We then say that the extraction $\mathscr{F}(\Psi)$ of the foliation from the wave function is covariant. If ${\mathscr{F}}$ is directly defined in terms of other covariant structures, it will itself be covariant. This will be the case when $\mathscr{F}$ 
is defined by a covariant vector field, such as 
$J^\Psi$  given by 
\eqref{tensors}  when it is integrable and time-like.

As another example, which is applicable to any relativistic quantum field theory, consider the total 4-momentum
\begin{equation}
P^\mu =  \int_S d\sigma_\nu(x)  \langle \Psi | t^{\mu \nu}(x) | \Psi \rangle,
\label{eq-4momentum}
\end{equation}
where $t^{\mu \nu}(x)$ is the energy-momentum tensor in the Heisenberg picture and $S$ is an arbitrary space-like hypersurface. By Stokes' theorem and the conservation equation 
\begin{equation}\label{conserv}
\partial_\mu t^{\mu \nu}(x)  = 0, 
\end{equation}
the four-vector $P^\mu$ does not depend on the choice of the
hypersurface $S$, so it defines a constant vector
field on space-time \cite{schweber61}. This vector field is generically
future-causal. (For the ground state, which is Lorentz invariant, it
will be zero. For massless particles it might be light-like.
But we set aside such unrealistic special cases.)
As such, it generically defines a foliation, of hyperplanes orthogonal to
it. The reference frame in which these hyperplanes correspond to
equal-time hyperplanes is the frame in which the total three-momentum
vanishes.

Finally, considering the Bohmian guidance law for Dirac particles, we see that the velocity field given in \eqref{EQU1} involves as geometrical structures only the multi-tensor $J$ given in \eqref{multicurrent}, which is covariant, and a foliation $\mathscr{F}$. For a covariant choice of foliation, such as the one determined by the vector field \eqref{eq-4momentum}, the guidance law is covariant.

\section{Generalization of the Bohmian particle dynamics}
\label{sec5}

A main objection to theories such as Bohmian mechanics is that they must inevitably conflict with Lorentz invariance because of their strong non-locality. In this section we  point out that, even when it does not define a foliation, the most familiar kind of space-time structure in quantum field theory -- a vector field on space-time -- may afford a reconciliation deemed widely impossible. For this it is entirely unnecessary that the exact quantum predictions emerge. It is enough that nonlocal effects arise in a completely covariant way.

We thus sketch here a possible relativistic generalization of the Bohmian dynamics that does not require a foliation, but merely a time-like vector field, and which reduces to the one with a foliation in the case the vector field is integrable. We only discuss this generalization for the case of a particle ontology. It is unclear whether it can be extended to a field ontology.

With $n$ a unit time-like vector field, the generalized law of motion reads, following \eqref{particleguidance},
\begin{equation} 
{\dot X}_k(x) \propto  v^{n, \Psi{}}_k (X_1^{\Sigma_x}, ... , X_N^{\Sigma_x})\,.
\end{equation}
Here $\Sigma_x$ is the surface through $x$ determined by the vector field $n$ in the manner described below. As usual, ${\dot X}_k(x)$ is the tangent vector to the $k^{\textrm{th}}$ particle's worldline at the point $x$ and $v^{n,\Psi}$ is a function on the Cartesian product of $N$ copies of space-time. The configuration $(X_1^{\Sigma_x}, ... , X_N^{\Sigma_x})$ at which the velocity field is evaluated is the collection of crossings by the particle world lines of the surface $\Sigma_x$. 

$\Sigma_x$ is the surface that is swept out by the curves through $x$ whose tangent vector at each point is normal to $n$ at that point and which are otherwise as straight as possible, so that their tangent vector only changes in the direction of $n$. Explicitly, in Hamiltonian form, the curves that generate the surface $\Sigma_x$ satisfy the equations
\begin{equation} 
\dot{x}^\mu = v^\mu , \qquad {\dot v}^\mu = - n^\mu v^\nu  v^\kappa \partial_\kappa n_\nu
\label{generalizedge}
\end{equation}
and have initial conditions $x(0)=x$ and $v(0) \cdot n(x(0))=0$ (i.e.\ the initial velocity is orthogonal to $n$). The dynamics then guarantees that the velocity remains orthogonal to $n$, i.e.\ that $v(\tau) \cdot n(x(\tau))=0$. (The dynamics \eqref{generalizedge} is the minimal modification of ``free motion'' in phase space that is compatible with the constraint $v \cdot n =0$.)

The surface $\Sigma_x$ itself need not be normal to $n$. It also need not be space-like. Only when $n$ is integrable must the surface $\Sigma_x$ be normal to $n$ and in that case $\Sigma_x$ is just one of the leaves of the foliation determined by $n$. The Bohmian dynamics then reduces to the one given before in \eqref{particleguidance}.

It is important to note that, unless $n$ is integrable, the surfaces $\Sigma_x$ do not form a foliation. Namely, while $x \in \Sigma_y \Leftrightarrow y \in \Sigma_x$, the surfaces $\Sigma_x$ and $\Sigma_y$ will generically be different. (This further implies that the relation $x \sim y \Leftrightarrow x \in \Sigma_y$, which is reflective and symmetric, fails to be transitive when $n$ is not integrable, and hence is not an equivalence relation.)

While this dynamics is a generalization of our earlier one in terms of a foliation, there is one drawback compared to that formulation. Namely, the relationship between its predictions and those of standard quantum theory is difficult to discern. This is an issue we turn to in the next section.

\section{Quantum equilibrium}
\label{sec6}
Non-relativistic Bohmian mechanics as outlined in the introduction reproduces the standard quantum predictions provided that the particle distribution for an ensemble of systems with the same wave function $\Psi(x)$ is given by $|\Psi(x)|^2$. This distribution is called the quantum equilibrium distribution. It satisfies the special property of equivariance: if the distribution is given by $|\Psi(x,t_0)|^2$ at a certain time $t_0$, then the Bohmian dynamics implies that the distribution is given by $|\Psi(x,t)|^2$ at all other times $t$.

In relativistic space-time, when the Bohmian dynamics is formulated using a foliation $\mathscr{F}$, there is again a distinguished distribution $\rho^\Psi$ (a generalized ``$|\Psi|^2$'' distribution) which plays the role of an equilibrium distribution and which is equivariant (i.e., has the same form as a functional of $\Psi$) along the leaves of the foliation. Since this distribution corresponds to the usual quantum distribution, the theory reproduces the standard quantum predictions \cite{duerr99}. For example, in the case of the Bohm-Dirac theory, on any leaf $\Sigma$ of $\mathscr{F}$, $\rho^\Psi$ corresponds to the distribution on $\Sigma^N$ given by
\begin{equation}
\rho^\Psi( x_1, \ldots, x_N) = J^{\mu_1 \ldots \mu_N} ( x_1, \ldots, x_N)  n^{\mathscr{F}}_{\mu_1}(x_1) \ldots   n^{\mathscr{F}}_{\mu_N}(x_N) .
\label{equilibrium}
\end{equation}
This is the distribution of the $N$ points of intersection of the world lines of the $N$ particles with $\Sigma$.

However, it has been shown that quantum non-locality implies that
quantum equilibrium can not hold in every reference frame, or with
respect to every foliation \cite{berndl96a}. That is, even though we
may assume equilibrium along the leaves of the foliation that was used
in formulating the Bohmian dynamics, the implied distribution along a
space-like hypersurface $\Sigma$ that is not a leaf of the foliation
may depend not only on $\Psi$, but also -- and in a complicated way
-- on the relationship between $\Sigma$ and the foliation. It may
therefore be different from the usual quantum
distribution. Nevertheless, as discussed in detail in
\cite{berndl96a,duerr99}, this will not change the fact that the
theory will reproduce the standard quantum predictions, assuming
quantum equilibrium along the leaves of the foliation. (This is closely related to Bell's remark that ``the laws of
    physics in any \emph{one} reference frame account for all physical
    phenomena, including the observations of moving observers'' \cite[p.\ 77]{bell76b}.) 

In the case of our generalized Bohmian dynamics of section \ref{sec5}, unless $n$ is integrable and determines a foliation, there is no distinguished distribution (or at least not an obvious one) that may play the role of an equivariant equilibrium distribution. (Recall that the surfaces $\Sigma_x$ that are used to formulate this generalized dynamics do not determine a foliation). In particular, there seems to be no reason to have the usual quantum distribution along the leaves of any foliation. As such, it seems unlikely that its predictions agree with those of standard quantum theory. (This is also true for the models in \cite{bell84,horton04,squires93,berndl96a,goldstein03,horton01,dewdney02,nikolic05b} that were mentioned in the introduction. These theories do not employ a distinguished foliation and hence it is hard to even begin a statistical analysis. In theories such as proposed in \cite{bell84,horton04}, the foliation depends on the actual configuration and different initial configurations will generically determine different foliations. Hence it is not clear how to compute the statistical predictions for ensembles of systems with the same initial wave function but different initial configurations of the local beables.) When $n$ is integrable, however, the dynamics reduces to the one in terms of a foliation, and a distribution like \eqref{equilibrium} is equivariant along the leaves of that foliation. The theory then reproduces the standard quantum predictions.

\section{Discussion}
\label{sec7}

We have developed here the idea that the privileged space-time
foliations (that have long been recognized as a crucial ingredient in Bohmian
theories) might not need to be separately posited as additional
space-time structure, but might instead be extracted from the objects
already present in the theory -- in particular, from the wave function
$\Psi$.  As explained in Section \ref{sec4}, there is a natural sense
in which this extraction can be understood as respecting the
appropriate space-time symmetries, so that a theory
formulated in this way should be regarded as fundamentally Lorentz
invariant.  

Several distinct possibilities for extracting a foliation from the
wave function (and one possibility for getting along without such a
foliation) have been suggested.  Further exploring these possibilities
is a natural subject for future work.  At present, rather than lobby
for any one particular such possibility, we wish to merely stress that
several quite workable options exist -- indeed, as explained
especially in Section \ref{sec3}, there would seem to be a plethora of
rather obvious candidates as soon as one considers quantum field
theories.

Everything in the type of theory proposed here
-- the dynamical law for $\Psi$, the rule for defining $\mathscr{F} =
\mathscr{F}(\Psi)$, and the guidance equation(s) for the local beables
-- is fully Lorentz covariant and thus seemingly entirely compatible
with the space-time structure being exclusively Minkowskian.  Is such
a theory then fundamentally -- and/or seriously -- relativistic?  

This is not an easy question to answer, because it is not at all clear
what, exactly, fundamental/serious compatibility with relativity does,
or should, require.  Lorentz invariance is clearly one necessary
requirement.  
But it is easy to imagine that someone might dismiss -- as clearly
\emph{incompatible} with relativity -- the type of theory
proposed here, simply on the grounds that it involves a dynamically
privileged foliation.  Frankly, it would be hard to disagree with this
sentiment.  But on the other hand, one of the important implications
of our proposal is that foliations can be extracted from
$\Psi$ and will therefore in a sense be present in \emph{any} kind of quantum
theory, Bohmian or otherwise.  So if the mere \emph{presence} of a 
foliation renders a theory un-relativistic, it seems hopeless that a
viable relativistic theory could ever be constructed.  

Other at-least-semi-reasonable criteria can be imagined.  For example,
some might feel that good relativistic theories should respect
relativistic local causality, or should posit exclusively local
beables, or should exhibit some kind of relativity principle for
sub-systems. Some of these have in fact been shown to be incompatible with experiment, and  we are not, at present, prepared to take any strong
position on their appropriateness, or that of other possible
criteria. Thus we are not in fact able to answer the question posed in this paper's title. We stress however that these criteria revolve around aspects of locality  that are largely  incompatible with quantum mechanics. Thus if Bohmian mechanics indeed \emph{cannot} be made relativistic, it seems likely that  quantum mechanics can't either.

\section{Acknowledgments}
We thank Roderich Tumulka for helpful discussions. This work is supported in part by COST (MP1006). N.Z.~is supported in part by INFN. S.G.~and W.S.~are supported in part by the John Templeton Foundation. The opinions expressed in this publication are those of the authors and do not necessarily reflect the views of the John Templeton Foundation.

\end{document}